\documentclass[11pt]{article}
\pdfoutput=1
\usepackage{jheppub}
\usepackage[OT1]{fontenc}
\usepackage{url}
\usepackage{bm}
\usepackage{graphicx}
\usepackage{epsfig}
\usepackage{rotating}
\usepackage{dcolumn}
\usepackage{xcolor}
\usepackage{slashed}
\usepackage{threeparttable}
\usepackage[normalem]{ulem} 
\usepackage{enumerate}

\title{\bf Equivalence between massless neutrinos and lepton number conservation in fermionic singlet extensions of the Standard Model}

\author{K. Moffat,}
\affiliation{Institute for Particle Physics Phenomenology, Department
  of Physics, Durham University, South Road, Durham DH1 3LE,
  United~Kingdom.}
\author{S. Pascoli,}

\author{C. Weiland}

\emailAdd{kristian.p.moffat@durham.ac.uk}
\emailAdd{silvia.pascoli@durham.ac.uk}
\emailAdd{cedric.weiland@durham.ac.uk}

\abstract{We discuss the most general necessary and sufficient condition for three massless light neutrinos in variants of the type I seesaw mechanism
in which we introduce an arbitrary number of fermionic gauge singlets. We find that having massless light neutrinos is equivalent to enforcing the conservation
of lepton number. As a consequence, any symmetry that leads to massless light neutrinos will contain as an unbroken subgroup a conserved lepton number. This
will be important for searches for heavy sterile neutrinos since in general the light neutrino masses will be proportional to small lepton number violating
parameters that will also suppress lepton number violating signatures.}

\preprint{IPPP/17/110}
\keywords{Beyond Standard Model, Neutrino Physics}

\allowdisplaybreaks[2]

\begin{document}

\thispagestyle{empty}
\def\thefootnote{\fnsymbol{footnote}}
\setcounter{footnote}{1}

\setcounter{page}{0}
\maketitle
\flushbottom

\def\thefootnote{\arabic{footnote}}
\setcounter{footnote}{0}

\section{Introduction}

The explanation of the observed neutrino oscillations~\cite{Olive:2016xmw} requires at least two neutrinos to have a non-zero mass.
However, the Standard Model (SM) in its original formulation cannot explain neutrino 
oscillations: lepton flavour as well as total lepton number is accidentally conserved and neutrinos are massless. Majorana masses are forbidden for the SM neutrinos at the renormalisable level although in an effective theory framework, at dimension five,  one may construct the Weinberg operator to generate these terms~\cite{Weinberg:1979sa}. 
One of the simplest neutrino mass generating mechanisms is the type
I seesaw~\cite{Minkowski:1977sc,Ramond:1979py,GellMann:1980vs,Yanagida:1979as,Mohapatra:1979ia,Schechter:1980gr,Schechter:1981cv} in which the Weinberg operator is UV-completed  by the addition of right-handed neutrinos to the SM and no
extra symmetry is enforced.

In high-scale seesaw models, these right-handed neutrinos are sub-dominant components of the light neutrino mass eigenstates whose lightness is proportional to the suppression of their sterile component.
Unfortunately, the very high mass of the heavy neutrinos and their small active component also
suppresses any phenomenological signature associated with the right-handed neutrinos.
In order to avoid this issue, many low-scale variants of the type I seesaw in which the new fermions are at the TeV scale or below have been 
proposed~\cite{Mohapatra:1986aw,Mohapatra:1986bd,Bernabeu:1987gr,Pilaftsis:1991ug,Ilakovac:1994kj,Akhmedov:1995ip,Akhmedov:1995vm,Barr:2003nn,Malinsky:2005bi,
Branco:1988ex,Buchmuller:1991tu,Asaka:2005an,Asaka:2005pn,Chao:2009ef,Chattopadhyay:2017zvs,Adhikari:2010yt,Dev:2012sg,Kang:2006sn,He:2009ua,Ibarra:2010xw,
Haba:2011pe,Ma:2009du,Wyler:1982dd,Chikashige:1980ui,Gelmini:1980re,Leung:1983ti,Petcov:1984nz} 
(see also~\cite{Chen:2011de,Boucenna:2014zba} and references therein)
which do not strongly suppress the active component of the heavy neutrinos. Instead, they rely on a cancellation between the contribution due to different sterile neutrinos to the light neutrino masses in order to increase the active-heavy neutrino mixing and
lower the seesaw scale. This leads to possible observable signatures of the heavy neutrinos at colliders or in low-energy experiments studying meson decays or lepton 
flavour violation.

Many searches for such low-scale seesaw models have focused on LNV signals since  they constitute striking signatures of a process that is forbidden in the SM and thus for which the background is suppressed~\cite{Olive:2016xmw,Aad:2015xaa,Khachatryan:2016olu,Aaij:2014aba,Shuve:2016muy,CERNNA48/2:2016tdo}.
Some low-scale seesaw models, for instance the inverse or linear seesaw, introduce an approximate lepton number symmetry in order to achieve the required cancellation. Unfortunately this approximate symmetry generally leads to a reduced rate for LNV signals. 
There are also models where an accidental cancellation is present at tree level, such as the Extended 
Seesaw~\cite{Kang:2006sn}. It should be emphasised that radiative corrections in these models may spoil these cancellations and lead to sizeable light neutrino masses so that LNV signatures are nevertheless suppressed.

It was shown by Kersten and Smirnov
that, in models with three or fewer fermionic gauge singlets of equal mass, requiring an exact radiatively stable cancellation of the first
term of the seesaw expansion is equivalent to requiring that lepton number is conserved~\cite{Kersten:2007vk}. This extends earlier results which did not consider the effects of radiative corrections~\cite{Ingelman:1993ve,Gluza:2002vs}. However, their result cannot be directly extended to models such as the inverse seesaw model where a larger number of gauge singlets is required to reproduce neutrino oscillation data. In this case, the condition derived from
the cancellation of the tree-level mass contributions is not sufficient any more. Besides, the requirement of equal masses is obtained via the running of the Weinberg operator when the Higgs boson is lighter than all heavy neutrinos. This motivates our work that proves that, for models with an arbitrary number
of sterile neutrinos that can also be lighter than the Higgs boson,
massless light neutrinos are equivalent to lepton number conservation  (LNC). This provides a firm basis to the requirement of a nearly conserved lepton number symmetry in low-scale seesaw models and implies that
any symmetry leading to massless light neutrinos contains lepton number as a subgroup or an accidental symmetry.

In section \ref{SingletNeutrinoExtensionsOfSM} we begin with a review of extensions of the SM with fermionic gauge singlets and fix our notations. In \ref{theorem} we expose our main result,
that requiring three massless light neutrinos is equivalent to a specific choice of the neutrino mass matrix that conserves lepton number independently of the number of heavy neutrinos. In \ref{Demo} we present a proof of this result. We conclude in section \ref{conclude} by discussing the implications of our result.

\section{Singlet neutrino extensions of the Standard Model} \label{SingletNeutrinoExtensionsOfSM}

We focus on extensions of the Standard Model that introduce $m$ gauge singlet neutrinos $N_{iR}$ for $i \in \{1,\hdots m\}$. After electroweak symmetry breaking the mass matrix reads
\begin{align}\label{Mdefinition}
\mathcal{L}_m & = -\frac{1}{2} \left(\bar{\nu}_L, \bar{N}^c_R \right) 
M
\left(\begin{array}{c}
\nu_L^c\\
N_R\\
\end{array}\right)
+\text{h.c.}\\
\nonumber & = -\frac{1}{2} \left(\bar{\nu}_L, \bar{N}^c_R \right) 
\left(\begin{array}{cc}
0 & m_D\\
m_D^T & m_R\\
\end{array}\right)
\left(\begin{array}{c}
\nu_L^c\\
N_R\\
\end{array}\right)
+\text{h.c.},
\end{align}
where $\nu_L=\left(\nu_{eL},\nu_{\mu L},\nu_{\tau L}\right)$ and $N_R=\left(N_{1R},\hdots,N_{mR}\right)$. Consequently the $\left(3 \times m\right)$ Dirac mass terms $m_D$ arise from the Higgs Yukawa coupling of the $SU(2)_L$ lepton doublets ($L_{e}$, $L_{\mu}$, $L_{\tau}$) with $N_R$. The gauge-invariant Majorana mass terms given by the $\left(m \times m \right)$ matrix $m_R$ violate lepton number by two units. For the rest of this article, we work with no loss of generality in
the basis in which $m_R$ is real and diagonal.

The unitary matrix $U$ performs the Takagi factorisation of the tree-level mass matrix through
\begin{equation} \label{Udefinition}
U^T M U = \hat{M},
\end{equation}
where a hat denotes a diagonal matrix.

The matrix $U$ may be factorised into two unitary matrices $U'$ and $U''$ such that
\begin{equation}
U=U'U''= U' \left(\begin{array}{cc}
U_l & 0\\
0 & U_h \\
\end{array}\right),
\end{equation}
where $U'$ performs the block-diagonalisation
\begin{equation}
U'^T M U' = 
\left(\begin{array}{cc}
M_{l} & 0\\
0 & M_{h} \\
\end{array}\right)
\end{equation}
where $M_{l}$ and $M_{h}$ are the light and heavy neutrino mass matrices respectively and the
matrix $U''$ acts to diagonalise the light and heavy mass matrices via $U_l$ and $U_h$ respectively:
\begin{equation}
\left( U' U'' \right)^T M \left( U' U'' \right) = 
\left(\begin{array}{cc}
\hat{M}_{l} & 0\\
0 & \hat{M}_{h} \\
\end{array}\right),
\end{equation}
where $\hat{M}_l=\text{diag}\left(m_1,m_2,m_3\right)$ and $\hat{M}_h=\text{diag}\left(m_4,\hdots,m_{m+3}\right)$.

It will be convenient to decompose $U$ into blocks as
\begin{equation}
U \equiv \left(\begin{array}{cc}
W & V\\
S & T \\
\end{array}\right).\end{equation}
The three light neutrino masses are given at tree-level by the singular values of $M_{l}$. At the first order in the seesaw expansion for the tree-level contributions, 
the light neutrino mass matrix is given by
\begin{equation}\label{lightmass}
M_{l} \approx -m_D m_R^{-1} m_D^T.
\end{equation}
This mass term is lepton number violating as can be expected since lepton number is not a symmetry of the model. Beyond tree-level this mass term receives radiative corrections from self-energy diagrams containing only the Higgs and Z bosons~\cite{Pilaftsis:1991ug,Grimus:2000vj,AristizabalSierra:2011mn}.

To first order in the seesaw expansion, we have that
\begin{equation} \label{heavylightmixing}
V \sim \mathcal{O} \left( \frac{m_D}{m_R} \right),
\end{equation}
which describes the small contribution of active neutrinos to the heavy neutrino states. A naive order of magnitude estimate of the light neutrino masses given in eq.(\ref{lightmass}) gives $m_R\sim10^{13}$~GeV for $m_D\sim100$~GeV if we require the light neutrino masses
to be around the eV scale.  However, this implies a very small $ V\sim \mathcal{O}\left(10^{-11}\right)$ in eq.(\ref{heavylightmixing}).
Experimental signatures of heavy nearly-sterile neutrinos are
strongly suppressed by this small mixing and also by their large masses.

There do however exist low-scale variants of the type I seesaw in which an approximate lepton number symmetry is introduced in order to prevent the strong suppression of the mixing parameters. 
Consider for instance the inverse seesaw in which the mass scale is lowered to the TeV or below making the heavy states accessible in direct searches. The model introduces two types of gauge-singlet neutrinos called $\nu_R$ and $X_L$. In the basis $\left(\nu_L, \nu_R^c, X_L \right)$ with lepton number
assignments $L\left(\nu_R\right)=+1$ and $L\left(X_L\right)=+1$, the mass matrix is then
\begin{equation}\label{ISS}
M_{\text{ISS}}=\left(\begin{array}{ccc}
0 & M_D^T & 0\\
M_D & \mu_R & M_R\\
0 & M_R^T &  \mu_X\\
\end{array}\right),
\end{equation}
in which $\mu_R$ and $\mu_X$ are small lepton number violating matrices.

In particular, for the inverse seesaw, from eq.(\ref{ISS}), assuming the seesaw limit $\mu_R,\mu_X \ll M_D \ll M_R$, the light masses are given at tree-level by
\begin{equation}
M_{l} \approx M_D^T M_R^{T-1} \mu_X M_R^{-1} M_D.
\end{equation}
The light neutrino masses deviate from zero in proportion to the lepton number violating parameter $\mu_X$. If the scale of $\mu_X$ is made small by an approximate lepton number symmetry, then the scale of $M_R$ need not be so large in order to generate a small light neutrino mass.
The mixing goes as $M_D^T M_R^{-1}$ and is thus not as suppressed as in the one generation case, leaving room for observable experimental signatures in these low scale seesaw variants.

These sorts of models suggest an interesting question: \textit{Can one have large lepton number violation and, at the same time, small masses for the light neutrinos?}

\subsection{Theorem}
\label{theorem}

Under the assumption that conditions i) and ii) (below) are obeyed,
the necessary and sufficient condition for three exactly massless neutrinos to all radiative orders when an arbitrary
number of gauge-singlet neutrino fields are added to the SM is that the neutrino mass matrix is given by
\begin{equation}\label{theoremresult}
\widetilde{M} =
\left(\begin{array}{cccc}
0 & \alpha & \pm i \alpha & 0\\
\alpha^T & M_1 & 0 & 0\\
\pm i \alpha^T & 0 & M_1 &  0\\
0 & 0 & 0 & M_2
\end{array}\right),
\end{equation}
in which $M_1$ and $M_2$ are diagonal matrices with positive entries and $\alpha$ is a generic complex matrix.  The conditions that must hold are:
\begin{enumerate}[i)]
\item there is no cancellation between different orders of the seesaw expansion. This is a necessary requirement to satisfy phenomenological constraints as mixing cannot be of order one
(see appendix~\ref{SeesawExpansionHierarchy}),
\item there are no fine-tuned cancellations between different radiative orders. These fine-tuned cancellations cannot be achieved solely by specific textures of the neutrino mass matrix (see appendix~\ref{finetuning}).
\end{enumerate}

The mass matrix of eq.(\ref{theoremresult}) may be related to those arising in the common low-scale seesaw
variants~\cite{Mohapatra:1986aw,Mohapatra:1986bd,Bernabeu:1987gr,Pilaftsis:1991ug,Ilakovac:1994kj,Akhmedov:1995ip,Akhmedov:1995vm,Barr:2003nn,Malinsky:2005bi,
Branco:1988ex,Buchmuller:1991tu,Asaka:2005an,Asaka:2005pn,Chao:2009ef,Chattopadhyay:2017zvs,Adhikari:2010yt,Dev:2012sg,Kang:2006sn,He:2009ua,Ibarra:2010xw,
Haba:2011pe,Ma:2009du,Wyler:1982dd,Chikashige:1980ui,Gelmini:1980re,Leung:1983ti,Petcov:1984nz,Chen:2011de,Boucenna:2014zba}.
Starting with the neutrino mass matrix $\widetilde{M}$, one can always find a unitary matrix
\begin{equation}
Q=
\left(\begin{array}{cccc}
1 & 0 & 0 & 0\\
0 & \pm\frac{i}{\sqrt{2}}D & \frac{1}{\sqrt{2}}D & 0\\
0 & \frac{1}{\sqrt{2}}D & \pm\frac{i}{\sqrt{2}}D & 0\\
0 & 0 & 0 & 1
\end{array}\right),
\end{equation}
with $D$ unitary. This may be used to change basis and perform a congruent transformation from the matrix of eq.(\ref{theoremresult}) to
\begin{align}
Q^T \widetilde{M} Q = 
\left(\begin{array}{cccc}
0 & \pm i \sqrt 2 (D^T \alpha^T)^T & 0 & 0\\
\pm i \sqrt 2 D^T \alpha^T & 0 & \pm iD^T M_1 D & 0\\
0 &  \pm iD^T M_1 D & 0 &  0\\
0 & 0 & 0 & M_2
\end{array}\right).
\end{align}
The latter is of the form
\begin{align}
\left(\begin{array}{cc}
M_\text{LNC} & 0\\
0 & M_2
\end{array}\right),
\end{align}
where 
\begin{equation}
M_\text{LNC}\equiv\left(\begin{array}{ccc}
0 & M_D^T & 0\\
M_D & 0 & M_R\\
0 & M_R^T &  0\\
\end{array}\right),
\end{equation}
in which we borrow the notation of eq.(\ref{ISS}).

Thus, there is a one-to-one correspondence between the lepton number conserving limit of the low-scale seesaw variants and the non-decoupled block of the mass matrix $\widetilde{M}$
of the theorem here presented.
The lepton number of the decoupled singlet neutrinos may be arbitrarily chosen without any phenomenological consequences, with zero leading
to a lepton number conserving model.
Therefore the theorem we are going to prove is equivalent to: \textit{The most general gauge-singlet neutrino extensions of the SM with no cancellation between different orders of the seesaw 
expansion, no fine-tuned cancellations between different radiative orders and which lead to three massless neutrinos are lepton number conserving.}

\subsection{Proof}\label{Demo}

The light neutrino masses receive contributions from both tree-level and radiative corrections and can be expanded in two convenient ways: \textit{i}) in the perturbative series 
in the couplings of the interaction Lagrangian giving radiative
corrections where each of these terms can be further expanded in \textit{ii}) the expansion in $m_D/m_R$ (the seesaw expansion). 

If one chooses to cancel terms in the radiative expansion with one another then one finds that an extreme fine-tuning is necessary~\cite{Lopez-Pavon:2015cga} (see appendix~\ref{finetuning}). We shall ignore such fine-tuned solutions and conclude that we must set the light masses to zero at tree-level,
then set them to zero at one-loop and so on.
It shall turn out to be necessary only to consider up to one-loop to achieve an all-orders massless result.

At each order of the perturbative expansion we disregard the possibility of having a cancellation 
between different orders of the seesaw expansion since it would lead to an active-heavy mixing larger than the experimental upper bounds (see appendix~\ref{SeesawExpansionHierarchy}). 
This problem does not occur if each term of the expansion is set to zero and we proceed to impose this condition in our proof.

\subsubsection{The matrix $\widetilde{M}$ as a sufficient condition for massless light neutrinos: \\ $\widetilde{M} \implies \hat{M}_l=0$}

The matrix $\widetilde{M}$ automatically leads to $M_l=0$ due to conservation of lepton number as demonstrated in section~\ref{theorem}. We provide an explicit proof at tree-level below. 

Consider the first term of the seesaw expansion at tree-level for the light neutrinos using the mass matrix of eq.(\ref{theoremresult}). Here we have,
\begin{equation}
m_D = \left( \alpha, \pm i \alpha, 0 \right),
\end{equation}
and
\begin{equation}
m_R^{-1} = 
\left(\begin{array}{ccc}
M_{1}^{-1} & 0 & 0\\
0 & M_{1}^{-1} & 0\\
0 & 0 & M_2^{-1}\\
\end{array}\right).
\end{equation}
Thus, the tree-level light mass at first order is
\begin{align}
m_D m_R^{-1} m_D^T & =
\left( \alpha, \pm i \alpha, 0 \right)
\left(\begin{array}{ccc}
M_{1}^{-1} & 0 & 0\\
0 & M_{1}^{-1} & 0\\
0 & 0 & M_2^{-1}\\
\end{array}\right)
\left(\begin{array}{c}
\alpha^T\\
\pm i \alpha^T\\
0\\
\end{array}\right)\\
& = \alpha M_1^{-1} \alpha^T + \left(\pm i\right)^{2} \alpha M_1^{-1} \alpha^T \\
& = 0.
\end{align}
Therefore, from eq.(\ref{lightmass}) we have
\begin{equation}
M_{l} = 0,
\end{equation}
considering only the first term of the seesaw expansion. However it is well-known that this is sufficient to set the entire seesaw expansion to zero at tree-level~\cite{Grimus:2000vj,Korner:1992zk}.
Now lepton number conservation implies that this massless condition is maintained at all orders in the loop expansion. We conclude from this that the mass matrix of eq.(\ref{theoremresult}) leads to three massless neutrinos to all orders.

\subsubsection{The matrix $\widetilde{M}$ as a necessary condition for massless light neutrinos: \\ $\hat{M}_l = 0 \implies \widetilde{M}$}

From
\begin{equation}
\hat{M}_l = 0,
\end{equation}
and the fact that we may always perform the singular value decomposition of $M_l$ (that is, $U_l$ always exists), we have that
\begin{equation}
M_l=U_l^* \hat{M}_l U_l^{\dagger}=0.
\end{equation}
Thus by consideration only of the first order in both expansions we have the condition\footnote{Recall that we bar cancellations between different orders of the radiative and seesaw expansions.}
\begin{equation} \label{mDmRmD}
m_D m_R^{-1} m_D^T=0.
\end{equation}

Following the condition in eq.(\ref{mDmRmD})~\cite{Grimus:2000vj,Korner:1992zk}, we define
\begin{equation} \label{Zdef}
Z=m_R^{-1} m_D^T
\end{equation}
and take
\begin{equation} \label{Ublock}
U'=
\left(\begin{array}{cc}
\left(1 + Z^{\dagger} Z \right)^{-\frac{1}{2}} & Z^{\dagger} \left(1+Z Z^{\dagger}\right)^{-\frac{1}{2}}\\
- \left(1+Z Z^{\dagger}\right)^{-\frac{1}{2}} Z & \left(1+Z Z^{\dagger} \right)^{-\frac{1}{2}} \\
\end{array}\right),
\end{equation}
which is unitary and block-diagonalises $M_{l}$ provided that eq.(\ref{mDmRmD}) holds.

With this notation, we find that
\begin{align}\label{lightmass2}
M_{l} = -\left(1+ Z^T Z^*\right)^{-\frac{1}{2}} m_D Z \left( 1 + Z^{\dagger} Z \right)^{-\frac{1}{2}},
\end{align}
where the presence of $m_D Z = m_D m_R^{-1} m_D^T = 0$ ensures that the entire seesaw expansion is zero. Thus at tree-level requiring the first term of the seesaw expansion
to cancel is enough to obtain $\hat{M}_l=0$.

Let us now consider the one-loop contribution to $\hat{M}_l$. We computed the one-loop induced mass for neutrinos and found it to agree with~\cite{Pilaftsis:1991ug}, giving 
\begin{equation} \label{selfenergy}
\delta M_{ij} = \Re \left[\frac{\alpha_W}{16 \pi^2 m_W^2} C_{ik} C_{jk} f\left(m_k\right)  \right],
\end{equation}
with $i,j \in \{ 1, 2, 3 \}$, $\alpha_W = g^2/4 \pi$ with $g$ the $SU(2)_L$ coupling constant,
\begin{equation}
C \equiv U_L^T U_L^*,
\end{equation}
\begin{equation}
U_L \equiv \left(W,V\right),
\end{equation}
and 
\begin{equation} \label{fdef}
f\left(m_k\right) = m_k \left(3 m_Z^2 g_{kZ} + m_H^2 g_{kH} \right)
\end{equation}
where
\begin{equation}
g_{ab} = \frac{m_a^2}{m_a^2-m_b^2} \log \frac{m_a^2}{m_b^2}.
\end{equation}
As $f\left(0\right)=0$, it is useful to define a diagonal matrix
\begin{align}
 F & \equiv \text{diag}(f(m_1),...,f(m_{m+3})), \\ & = \left(\begin{array}{cc}
0 & 0\\
0 & F_{h} \\
\end{array}\right),
\end{align}
such that eq.(\ref{selfenergy}) may be rewritten in matrix form as 
\begin{equation}
\delta M_{ij} = \Re \left[\frac{\alpha_W}{16 \pi^2 m_W^2} C F C^T \right]_{ij}.
\end{equation}

Imposing zero masses for the light neutrinos implies that the total one-loop self-energy must be set to zero{\footnote{Massless neutrinos must have zero imaginary parts for their self energy as they cannot decay and thus they have zero total self-energy.}}. 
This implies that the $\left(1,1\right)$ block of $C F C^T=0$, that is
\begin{equation}
\left(C F C^T\right)_{11} =W^T V^* h V^{\dagger} W =0,
\end{equation}
which may equivalently be written as
\begin{equation}
U_{l}^T \left( 1 + Z^T Z^* \right)^{-1} Z^T U_{h}^* F_{h} U_{h}^{\dagger} Z \left( 1 + Z^{\dagger} Z \right)^{-1} U_{l} = 0.
\end{equation}
This reduces to 
\begin{equation} \label{simplifiedoneloopcondition}
Z^T U_{h}^* F_{h} U_{h}^{\dagger} Z = 0
\end{equation}
upon the left and right multiplication by 
\begin{equation*}
\left( 1 + Z^T Z^* \right) U_{l}^*
\end{equation*}
and
\begin{equation*}
U_{l}^{\dagger} \left( 1 + Z^{\dagger} Z \right)
\end{equation*}
respectively.

Since $m_R$ is diagonal and positive, we have to the first order in the seesaw expansion
\begin{equation}
U_{h} \approx 1.
\end{equation}
Thus, again treating the terms of the seesaw expansion independently, from eq.(\ref{simplifiedoneloopcondition}) we arrive at
\begin{equation} \label{OLCondition}
Z^T F_{h} Z = 0,
\end{equation}
from the first term.

We shall now consider the implication of eq.(\ref{OLCondition}) for the form of the neutrino mass matrix and prove that it leads to eq.(\ref{theoremresult}). We use the tree-level expression for $Z$. Allowing for degenerate masses in $m_R$, in
the flavour-basis the mass matrix can be written
\begin{equation} \label{notationmassmatrix}
M=\left(\begin{array}{ccccc}
0 & m_{D1} & m_{D2} & \hdots & m_{Dn} \\
m_{D1}^T & \tilde{\mu}_1 & 0 & \hdots & 0\\
m_{D2}^T & 0 & \tilde{\mu}_2 & \hdots & \vdots\\
\vdots & \vdots & \vdots & \ddots & 0\\
m_{Dn}^T & 0 & \hdots & 0 & \tilde{\mu}_n \\
\end{array}\right),
\end{equation}
where each block $\tilde{\mu}_i$ is proportional to an $n_i \times n_i$ identity matrix with a mass $\mu_i$, ($\tilde{\mu}_i = \mu_i I_{n_i}$).
Correspondingly, we may write
\begin{align}
Z & = m_R^{-1} m_D^T\\
& = 
\left(\begin{array}{c}
\mu_1^{-1} m_{D1}^T\\
\mu_2^{-1} m_{D2}^T\\
\vdots \\
\mu_n^{-1} m_{Dn}^T\\
\end{array}\right).
\end{align}
In this notation eq.(\ref{OLCondition}) becomes
\begin{align} \label{summationequation}
Z^T  F_h Z & = \sum_{i=1}^n \mu_{i}^{-2} m_{Di} m_{Di}^T f\left(\mu_i\right) = 0.
\end{align}

Now, if the texture of the neutrino mass matrix is to determine the condition for massless neutrinos, an overall scaling
\begin{align} \label{massmatrixscaling}
M
\rightarrow
\Lambda M
\end{align}
does not affect the form of the mass matrix or the condition\footnote{Such a scaling removes the possibility of fine-tuned solutions in which a particular numerical choice of entries (in given units) for the mass matrix may lead to a cancellation. We attempt to quantify the degree of fine-tuning in appendix \ref{finetuning}.}  $\hat{M}_{l}=0$.
We shall show that this scaling leads to the condition
\begin{equation}
m_{Di} m_{Di}^T = 0.
\end{equation}
In fact the above scaling implies
\begin{equation}
 U^* \hat{M} U^\dagger \rightarrow \Lambda U^* \hat{M} U^\dagger = U^* \Lambda \hat{M} U^\dagger,
\end{equation}
and since $U$ is unitary by construction it cannot be redefined to absorb the scaling. As a consequence, the scaling promotes
\begin{equation} \label{Scaling}
f \left( m_i \right) \rightarrow f \left( \Lambda m_i \right)
\end{equation}
and, in the limit of the first term of the seesaw expansion in which $\hat{M}_h=m_R$
\begin{equation}
f \left( \mu_i \right) \rightarrow f \left( \Lambda \mu_i \right).
\end{equation}

We notice that  $f$ is a monotonically increasing strictly convex function, as shown in appendix \ref{f_function}. Thus one may choose $k > n$ (with $n$ defined in eq.(\ref{notationmassmatrix})) distinct values
for $\Lambda$ and obtain as many distinct equations of the form
\begin{equation}
 \sum_{i=1}^n \mu_{i}^{-2} m_{Di} m_{Di}^T f\left(\Lambda \mu_i\right) \Lambda^{-2} = 0
\end{equation}
These equations form a system of linearly independent equations for the coefficients 
\begin{equation*}
\mu_{i}^{-2} m_{Di} m_{Di}^T f\left(\mu_i\right).
\end{equation*} 
Since none of the $\mu_{i}$ are zero by construction,
the only solution of this system of equations is
\begin{equation} \label{mDmDT}
m_{Di} m_{Di}^T = 0.
\end{equation}

We shall now see that the condition of eq.(\ref{mDmDT}) is equivalent to having the neutrino mass take the form of eq.(\ref{theoremresult}). First, we
express each $m_{Di}$ in terms of vectors $u^i$, $v^i$, $w^i$ as
\begin{equation}
m_{Di}^T = \left( u^i, v^i, w^i \right).
\end{equation}
Then, we have
\begin{align}
m_{Di} m_{Di}^{T} =  \left(\begin{array}{ccc}
u^{iT} u^i & u^{iT} v^i & u^{iT} w^i\\
v^{iT} u^i & v^{iT} v^i & v^{iT} w^i\\
w^{iT} u^i & w^{iT} v^i & w^{iT} w^i
\end{array}\right)
\end{align}
and
\begin{align} \label{dmconditions}
u^{iT}u^i&=0 \\
\nonumber v^{iT}v^i&=0 \\
\nonumber w^{iT}w^i&=0 \\
\nonumber u^{iT}v^i&=0 \\
\nonumber u^{iT}w^i&=0 \\
\nonumber w^{iT}v^i&=0.
\end{align}
From a vector $u^i$ such that $u^{iT} u^{i}=0$, it is always possible to construct an orthogonal block-diagonal matrix $R_u=\text{diag}\left(R_u^1,\hdots,R_u^n \right)$ (appendix \ref{OrthoTrans}) such that
\begin{align}
\left( u^i, v^i, w^i \right) & \rightarrow \left( u^{i '}, v^{i '}, w^{i '} \right) \\
& = \left(R^i_u u^i, R^i_u v^i, R^i_u w^i \right),
\end{align}
in which 
\begin{equation}\label{uprime}
u^{i'}=\left(u^{i'}_1,\pm i u^{i'}_1, 0, \hdots, 0\right)^T.
\end{equation}
As a special case, if the original vector $u^{i}$ has only real components, then $u^{i'}=0$.
Such a transformation leaves $m_R$ unaffected as
\begin{align}
m_R'&  = \text{diag}\left( R_u^1 \tilde{\mu}_1 R_u^{1T}, \hdots, R_u^n \tilde{\mu}_n R_u^{nT} \right)\nonumber \\
& = \text{diag}\left( \tilde{\mu}_1, \hdots, \tilde{\mu}_n \right) \nonumber \\
& = m_R.
\end{align}
Under this transformation, we have
\begin{equation}
u^{iT}v^i=0 \rightarrow u^{i'T}v^{i'}=0,
\end{equation}
leading us to conclude that
\begin{equation}
v^{i'}=\left(v^{i'}_1,\pm i v^{i'}_1, v^{i'}_3, v^{i'}_4, \hdots, v^{i'}_{n_i} \right)^T.
\end{equation}
Similarly, we construct a second matrix $R_v$ acting on $\left(v^{i'}_3, v^{i'}_4, \hdots, v^{i'}_{n_i} \right)^T$ such that $v^{i'}$ is reduced to
\begin{equation}
v^{i''}=\left(v^{i'}_1,\pm i v^{i'}_1, v^{i''}_3, \pm i v^{i''}_3, 0 , \hdots, 0\right)^T.
\end{equation}
Finally, this process is repeated with $R_w$ such that 
\begin{equation}
w^{i'''}=\left(w^{i'}_1,\pm i w^{i'}_1, w^{i''}_3, \pm i w^{i''}_3, w^{i'''}_5 , \pm i w^{i'''}_5 \hdots, 0\right)^T.
\end{equation}
Each block of $m_D$ thus takes the form
\begin{equation}\label{ReducedDiracMass}
m_{Di} =
\left( \begin{array}{ccccccccc}
u^{i'}_1 & \pm iu^{i'}_1 & 0 & 0 & 0& 0 & 0 & \hdots & 0 \\
v^{i'}_1 & \pm iv^{i'}_1 & v^{i''}_3 & \pm i v^{i''}_3 & 0 & 0 & 0 & \hdots & 0 \\
w^{i'}_1 & \pm i w^{i'}_1 & w^{i''}_3 & \pm i w^{i''}_3 & w^{i'''}_5 & \pm i w^{i'''}_5 & 0 & \hdots & 0 \\
\end{array}\right).
\end{equation}
By rearranging the columns and rows, we may write the flavour-basis mass matrix as
\begin{equation}\label{endresult}
M = 
\left(\begin{array}{cccc}
0 & \alpha & \pm i \alpha & 0\\
\alpha^T & M_1 & 0 & 0\\
\pm i \alpha^T & 0 & M_1 &  0\\
0 & 0 & 0 & M_2
\end{array}\right) = \widetilde{M},
\end{equation}
where $\alpha$ are blocks constructed from a permutation of the columns of $m_D$ and $M_1$ and $M_2$ are positive diagonal matrices made from the same permutation of the diagonal entries of $\mu_i$.

We conclude that this neutrino mass matrix appears in any extensions of the Standard Model which introduce only new fermionic gauge singlets and in which the three light neutrinos are exactly massless (subject to the conditions previously discussed).

\section{Conclusions} \label{conclude}
In this article we have shown that requiring all three light neutrinos to be massless at all orders in perturbation theory is equivalent to choosing the neutrino mass matrix according to
eq.(\ref{theoremresult}). As a corollary, we found that this is equivalent to requiring lepton number to be conserved. This extends and generalizes a previous result by Kersten and Smirnov which was limited to
three heavy neutrinos or fewer with equal masses. This is particularly important since it provides a firm basis to the requirement of a nearly conserved lepton number symmetry in low-scale seesaw models. It also implies that
any symmetry leading to massless light neutrinos contains lepton number as a subgroup or an accidental symmetry.

However, neutrino oscillations imply that at least two of the three light neutrinos are not massless. This is only possible if lepton number is not conserved and indeed
many low-scale seesaw models relate the smallness of the light neutrino masses to the size of lepton number violation. This raises the question of the observability of the heavy neutrino 
contributions to LNV processes since we expect their contribution to be either suppressed by a small active-heavy mixing and large heavy neutrino masses or by the nearly conserved lepton number symmetry that are required to 
generate small enough masses for the light neutrinos. While this question was already addressed
in~\cite{LopezPavon:2012zg,Pascoli:2013fiz,Haba:2016lxc,Lopez-Pavon:2015cga,Hernandez:2016kel,Drewes:2016lqo} for neutrinoless double beta decay, we defer the study
of the collider implications to a subsequent article. 

\section*{Acknowledgements}

The authors receive financial support from the European Research Council under the European Union’s Seventh Framework Programme (FP/2007-2013)/ERC Grant
NuMass Agreement No. 617143. S.P. would also like to acknowledge partial support from the European Union’s Horizon 2020 research and innovation programme
under the Marie Sklodowska-Curie grant agreements No 690575 (RISE InvisiblesPlus) and No 674896 (ITN ELUSIVE), from STFC and from the Wolfson Foundation
and the Royal Society, and also thanks IFT-UAM/CSIC and SISSA for support and hospitality during part of this work.

\appendix

\section{Construction of $R^i_u$, $R^i_v$, $R^i_w$}\label{OrthoTrans}
We provide here a procedure for explicitly constructing the matrices $R^i_u$, $R^i_v$ and $R^i_w$ for given vectors $u^i$, $v^i$ and $w^i$.
We first construct
\begin{equation}
Y^i = u^{i*} u^{iT} + u^{i} u^{i \dagger}.
\end{equation}
As this is a real symmetric matrix, it is possible to choose a set of $n_i$ real orthogonal eigenvectors $b^i$. Then
\begin{equation}
R^i_u = \left(\begin{array}{cccc}
b^{iT}_1 \\
b^{iT}_2 \\
\vdots \\
b^{iT}_{n_i} \\
\end{array}\right)
\end{equation}
will perform the required transformation for generic $u^i$.

From the relations
\begin{equation}
\text{rank} \left( Y^i \right) \leq  \text{rank} \left( u^{i*} u^{iT} \right) + \text{rank}\left( u^{i} u^{i \dagger} \right)
\end{equation}
and
\begin{equation}
\text{rank} \left( u^{i*} u^{iT} \right) = \text{rank} \left( u^{i} u^{i \dagger} \right) = 0 \text{ or } 1,
\end{equation}
it follows that the rank of $Y^i$ may be at most $2$.
If it is rank $0$ then $u^i=0$ and this is a trivial case. If it is rank $1$ then $u^i$ is a real vector and cannot achieve $u^{iT} u^i = 0$ unless $u^i=0$ in this case.
Thus, for non-trivial $u^i$, $Y^i$ has rank $2$. Consequently it has $n_i-2$ eigenvalues equal to zero.

If $b^i_k$ corresponds to eigenvalue zero then $Y^i b^i_k = 0$. Thus,
\begin{equation}
u^{i} \left( u^{i\dagger} b^i_k \right) + u^{i*} \left( u^{iT} b^i_k \right) = 0.
\end{equation}
Multiplying on the left by $u^T$ yields
\begin{equation}
||u^{i}||^2 \left( u^{iT} b^i_k \right) = 0.
\end{equation}
which implies $u^{iT} b^i_k = 0$ (excluding the trivial solution $u^i=0$).

Finally,
\begin{equation}
u^{i'}= R^i_u u^i = \left(\begin{array}{cccc}
b^{iT}_1 u^i \\
b^{iT}_2 u^i\\
\vdots \\
b^{iT}_{n_i} u^i
\end{array}\right)
\end{equation}
which has components all zero except for two. Taking these components to be $u^{i'}_1$ and $u^{i'}_2$, the condition $u^{i'T} u^{i'}=0$ is equivalent to
\begin{equation}
 u^{i'2}_1 + u^{i'2}_2=0,
\end{equation}
which admits the solution
\begin{equation}
 u^{i'}_2=\pm i u^{i'}_1,
\end{equation}
and $u'$ thus takes the form of eq.(\ref{uprime}).

The matrices $R^i_2$ and $R^i_3$ can then be constructed by analogy. In the case of $R^{i}_{2}$ it is only necessary to repeat the above procedure with the vector 
\begin{equation*}
\left(v^{i'}_3, v^{i'}_4, \hdots \right)
\end{equation*}
 in place of $u^i$ from the start. This works as it gives zero upon taking its scalar product with itself. Similarly, $R^{i}_{3}$ is constructed by repetition of this argument with
\begin{equation*}
\left(w^{i''}_5, w^{i''}_6, \hdots \right)
\end{equation*}
in place of $u^i$.

\section{Properties of $f$}\label{f_function}

The function $f$
is composed of the sum of two terms of the form,
\begin{equation}
h(x) \equiv a \frac{x^3}{x^2-1}\log\left(x^2\right),
\end{equation}
for $x>0$. As $h$ is monotonic increasing and strictly convex then so is $f$. Since $a>0$ and is a constant, it will not affect the monotonicity and curvature of $h$ and we will drop it for the rest of this study.
We demonstrate that $h$ is monotonic increasing and strictly convex now.
\subsection{Monotonic increasing}
A change of variable $x \rightarrow e^u$ gives
\begin{equation}
h\left(u\right)= \frac{2 u}{e^{2u}-1} e^{3u},
\end{equation}
for $u \in \rm I\!R$. From this we obtain
\begin{equation}
h'\left(u\right) = e^{2 u} \text{csch}u \left(1-u (\coth u-2)\right).
\end{equation}
Since
\begin{equation}
h'\left(x\right)=u'\left(x\right) h'\left(u\right),
\end{equation}
and 
\begin{equation}
u'\left(x\right)= \frac{1}{x},
\end{equation}
which is strictly positive for $x >0$, studying the sign of $h'(x)$, requires us to concern
ourselves only with the sign of $h'(u)$. We may drop the factor $e^{2 u}$ and consider
\begin{equation*}
\text{csch}u \left(1-u (\coth u-2)\right),
\end{equation*}
where we recall that
\begin{equation}
\text{sgn}\left(\text{csch}u\right) = \text{sgn}\left(u\right).
\end{equation}

Starting with the result that
\begin{equation}
e^{-2u} \left(2u+1\right) < 1
\end{equation}
for strictly positive $u$, write
\begin{equation}
e^{-2u} \left(2u+1\right) - 1 < 0.
\end{equation}
Recognising the left-hand side as
\begin{equation}
2 u e^{-2u} -\left( 1 - e^{-2u}\right),
\end{equation}
we write
\begin{equation}
\left( 1 - e^{-2u}\right) \left( \frac{2 u e^{-2u}}{1-e^{-2u}} -1 \right) < 0,
\end{equation}
for strictly positive $u$.
Using the expression
\begin{equation}
2 e^{-2u} u = u \left(1 + e^{-2u} -1 + e^{-2u} \right)
\end{equation}
leads to the conclusion
\begin{equation}
u \coth u - 1 < u
\end{equation}
for strictly positive $u$.

Using the definition of $\coth$, we write
\begin{align}
 u \coth u -1 & = \frac{1 + e^{2u} (u-1) +u}{e^{2u} -1} \nonumber \\
	      & = \frac{e^{-u} (1+u) + e^u (u-1)}{e^u-e^{-u}}.
\end{align}
For $u>0$, we have $e^u - e^{-u} > 0$ and the sign of $u \coth u -1$ is given by the sign of
\begin{equation}
 \lambda (u) = e^{-u} (1+u) + e^u (u-1).
\end{equation}
Its derivative is 
\begin{equation}
 \lambda' (u) = u(e^{u} - e^{-u}),
\end{equation}
which is strictly positive for $u>0$. Thus $\lambda$ is a strictly increasing function on $\mathbb{R}^{+*}$ and its minimum on $\mathbb{R}^{+}$ is
\begin{equation}
 \lambda (0) = 0.
\end{equation}
From this, we have for $u>0$
\begin{equation}
u \coth u -1 > 0
\end{equation}
and since $u \coth u - 1$ is an even function of $u$, we also learn that 
\begin{equation}
u \coth u - 1 > u
\end{equation}
for $u<0$.

From this follows
\begin{equation}
u \coth u - 1 - 2u < 0
\end{equation}
for strictly positive $u$ and
\begin{equation}
u \coth u - 1 - 2u > 0
\end{equation}
for strictly negative $u$. We can also evaluate 
\begin{equation}
\lim_{u\rightarrow 0} h'(u) = 2.
\end{equation}

Thus $h'\left(u\right)>0$, the function $h\left(u\right)$ is monotonic increasing and so is $f\left(x\right)$.

\subsection{Strictly convex}
The strict convexity of $h$ may be demonstrated by considering the sign of its second derivative which may be expressed as
\begin{align}
\frac{d^2 h}{dx^2} = \frac{1}{x^2} \left( \frac{d^2 h}{du^2} -\frac{d h}{d u}\right).
\end{align}
The content of the parentheses written explicitly as a function of $u$ is
\begin{equation}
\frac{2 e^{3 u} \left(2 e^{2 u} (u-3)+e^{4 u}+6 u+5\right)}{\left(e^{2 u}-1\right)^3}.
\end{equation}
The denominator of this expression has sign equal to the sign of $u$. Our strategy for proving the convexity is to prove that this same statement may be made about the numerator. 

Owing to the positivity of $e^{3u}$, we need only consider now the sign of
\begin{equation}
s\left(u\right) = 2 e^{2 u} (u-3)+e^{4 u}+6 u+5.
\end{equation}
Let us observe that at $u=0$, 
\begin{equation}
s\left(0\right)=0,
\end{equation}
\begin{equation}
s'\left(0\right) = 0,
\end{equation}
\begin{equation}
s''\left(0\right) = 0,
\end{equation}
where the primes denote differentiation with respect to $u$. 
Consider now
\begin{equation}
s''\left(u\right) = 8 e^{2 u} u-16 e^{2 u}+16 e^{4 u}.
\end{equation}
As $e^{4u}/e^{2u}=e^{2u}$, we see that $s''(u)$ is positive for $u>0$ (since $e^{2u}>1$) and $s''(u)$ is negative for $u<0$ (since $e^{2u}<1$).

This implies that $s'\left(u\right)$ decreases when $u<0$ to the value $0$ at $u=0$ and increases for all positive $u$. That is to say that $s'\left(u\right)>0$ for all $u \neq 0$.

In turn, this implies that $s\left(u\right)$ is an increasing function for all negative and positive values of $u$. As $s\left(0\right)=0$,
then for $u<0$ we have $s\left(u\right)<0$ and for $u>0$ we have $s\left(u\right)>0$.

Therefore
\begin{equation}
\frac{2 e^{3 u} \left(2 e^{2 u} (u-3)+e^{4 u}+6 u+5\right)}{\left(e^{2 u}-1\right)^3} > 0
\end{equation}
for all $u \neq 0$. Besides, 
\begin{equation}
 \lim_{u\rightarrow 0} \frac{2 e^{3 u} \left(2 e^{2 u} (u-3)+e^{4 u}+6 u+5\right)}{\left(e^{2 u}-1\right)^3} = \frac{5}{3},
\end{equation}
and $s$ is always positive. We conclude that $h''\left(x\right)>0$ for all positive $x$.

\section{The cancellation of terms in the seesaw expansion} \label{SeesawExpansionHierarchy}
Alternatives to the condition of eq.(\ref{mDmRmD}) for the tree-level mass involve the cancellation of terms in the seesaw expansion. Consider the light mass matrix to second order in the expansion (denoted $M^{(2)}_{l}$),
\begin{equation}
M^{(2)}_{l} = - M^{(1)}_{l} + \frac{1}{2}  \left( M^{(1)}_{l} Z^{\dagger} Z + Z^T Z^{*} M^{(1)}_{l} \right),
\end{equation}
with $M^{(1)}_{l}$ the first order expression.

If this is set to zero by a cancellation of the two terms (as opposed to setting $M^{(1)}_{l}=0$), one finds that
\begin{equation}
0 = - \hat{M}^{(1)}_{l} + \frac{1}{2}  \left( \hat{M}^{(1)}_{l} \theta + \theta^T \hat{M}^{(1)}_{l} \right),
\end{equation}
where $\theta$ is $Z^{\dagger} Z$ transformed under a unitary transformation.

From the diagonal elements one finds
\begin{equation}
-\hat{M}^{(1)}_{lii} + \hat{M}^{(1)}_{lii} \theta_{ii}
\end{equation}
with no summation implied ($i \in \{ 1 , 2 ,3 \}$). Thus if one wants to avoid the solution that all three $\hat{M}^{(1)}_{lii}=0$, then it follows that at least one $\theta_{ii}=1$.

The Frobenius norm of a matrix $\theta$ is defined by
\begin{equation}
||\theta||_\text{F} = \sqrt{ \sum_{i=1}^3 \sum_{j=1}^3 |\theta_{ij}|^2 } = \sqrt{\text{Tr}\left( \theta \theta^{\dagger} \right)} .
\end{equation}
Now, $Z^{\dagger} Z$ and $\theta$ differ only by a unitary transformation and thus have the same Frobenius norm.

Using the 2$\sigma$ upper bounds on $Z^{\dagger}Z$ from the global fit\cite{Fernandez-Martinez:2016lgt}, we find $||\theta||_\text{F} \leq 0.0075$.
But the matrices resulting from the cancellation of the first pair of terms in the seesaw expansion have $||\theta||_\text{F} \geq 1$. This naturally precludes the possibility of having a 
cancellation between different orders of the seesaw expansion.

\section{Fine-tuning of the cancellation between the tree-level and one-loop contributions to the light neutrino masses}\label{finetuning}
An explicit caveat to our result is the possibility that the smallness of the light neutrino masses results from a cancellation between large tree-level and one-loop contributions as presented 
in~\cite{Lopez-Pavon:2015cga}. We will not discuss the radiative stability of this result. Instead we will show that this type of cancellation does not result from 
the texture of the neutrino mass matrix but from an extremely fine-tuned adjustment of all parameters, including the heavy neutrino masses.

\begin{figure*}[t]
\centering
\includegraphics[width=0.49\textwidth]{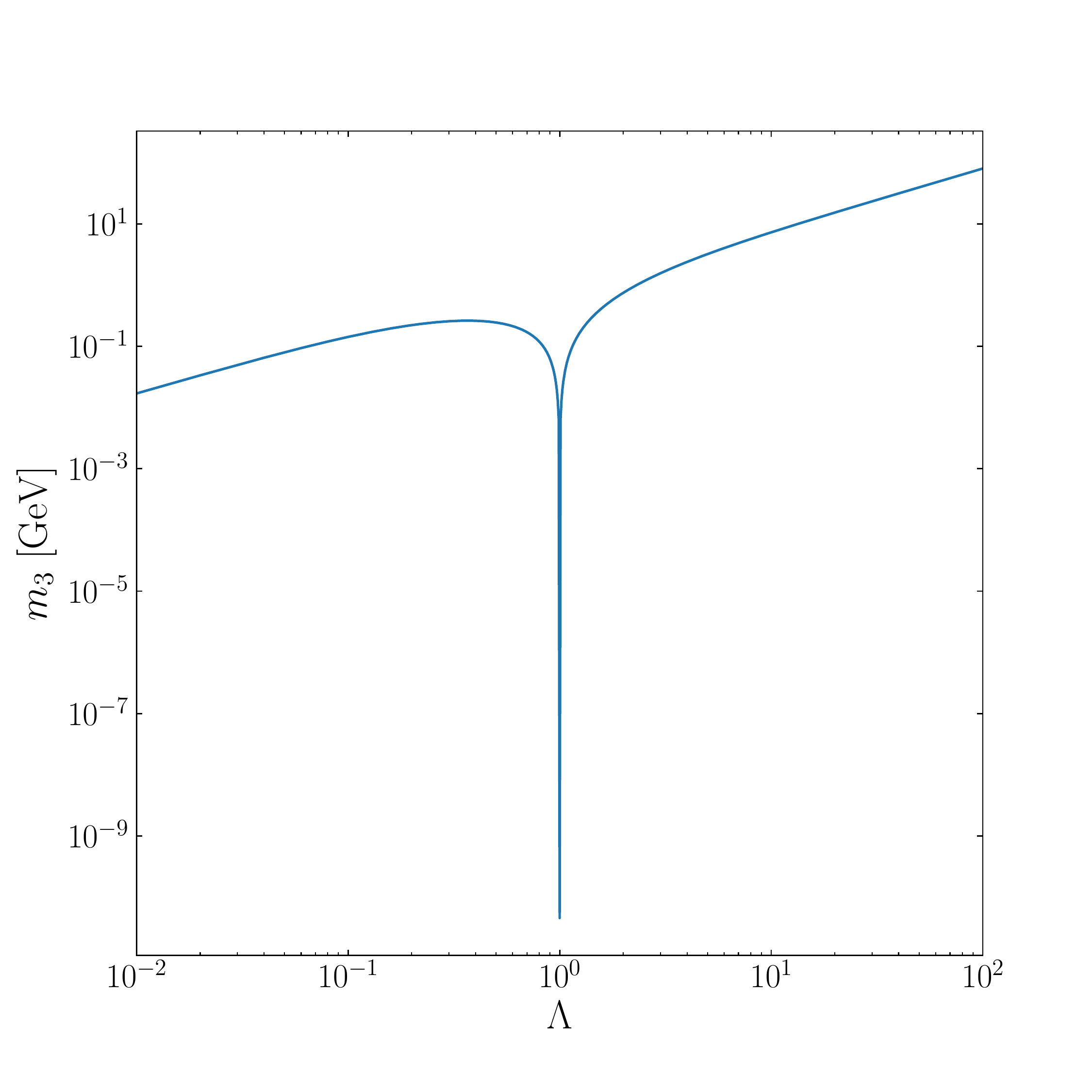}
\hfill
\includegraphics[width=0.49\textwidth]{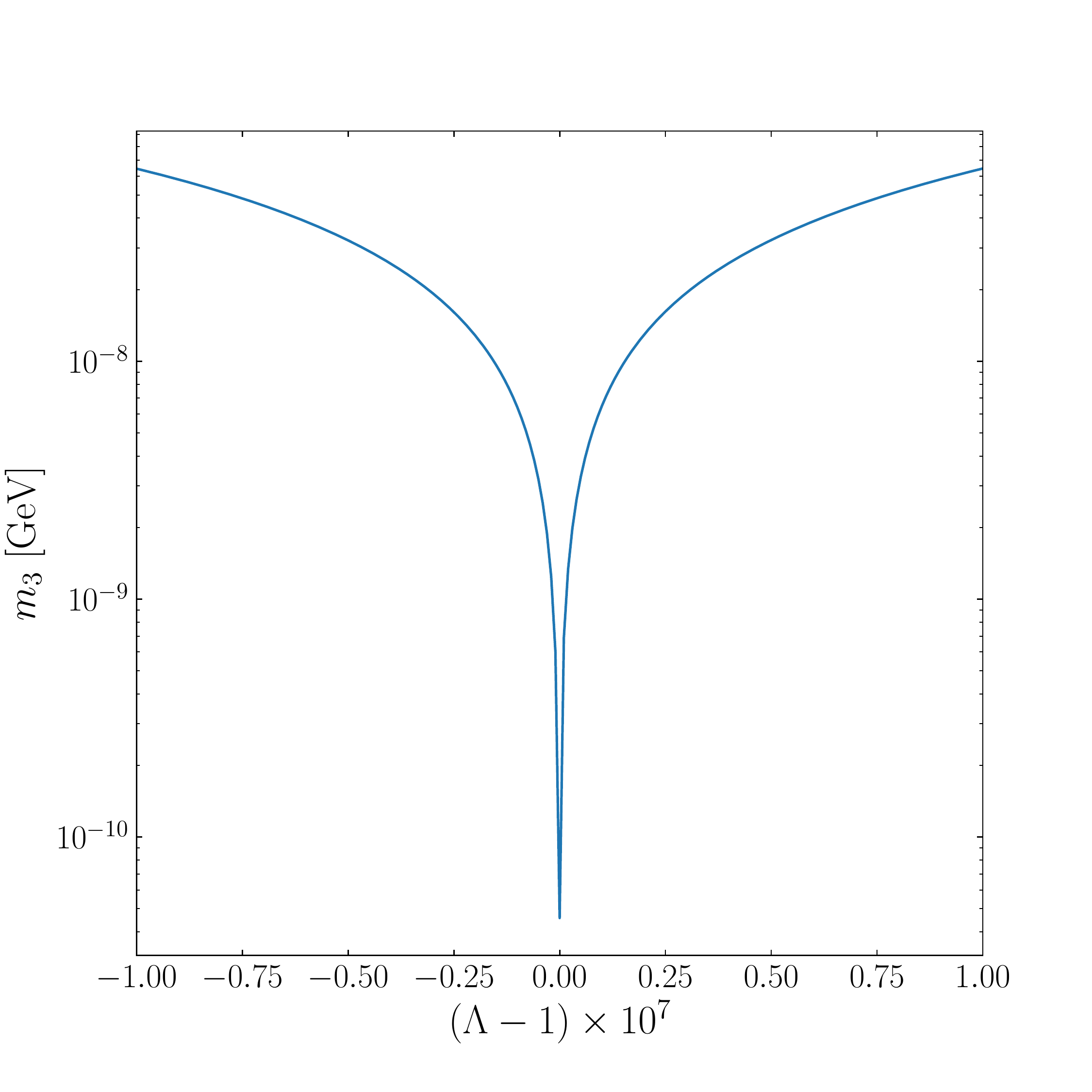}
\caption{Evolution of the mass ($m_3$) of the heaviest of the light neutrinos as a function of the rescaling parameter $\Lambda$. Input masses and couplings where chosen to give 
$m_\nu=m_\text{tree}+m_\text{1-loop}=0.046\,\text{eV}$ at $\Lambda=1$.}\label{PlotLambda}
\end{figure*}
Using the scaling introduced in eq.(\ref{massmatrixscaling}),
we plot in figure~\ref{PlotLambda} the evolution of the mass of the heaviest of the light neutrinos as a function of the rescaling parameter $\Lambda$. It is clear that even an extremely small 
deviation from $\Lambda=1$, less than $10^{-7}$ here, is enough to spoil the cancellation and lead to light neutrino masses in contradiction with experimental limits from $\beta$
decay~\cite{Kraus:2004zw,Aseev:2011dq} and observational cosmology~\cite{Ade:2015xua}. This demonstrates that such a cancellation cannot be achieved solely by the choice of a specific texture for
the neutrino mass matrix but relies on an extremely fine-tuned choice of the input masses.

\bibliography{MPW_LNC}{}

\providecommand{\href}[2]{#2}\begingroup\raggedright\begin{thebibliography}{10}

\bibitem{Olive:2016xmw}
{\scshape Particle Data Group} collaboration, C.~Patrignani et~al.,
  \emph{{Review of Particle Physics}},
  \href{https://doi.org/10.1088/1674-1137/40/10/100001}{\emph{Chin. Phys.}
  {\bfseries C40} (2016) 100001}.

\bibitem{Weinberg:1979sa}
S.~Weinberg, \emph{{Baryon and Lepton Nonconserving Processes}},
  \href{https://doi.org/10.1103/PhysRevLett.43.1566}{\emph{Phys. Rev. Lett.}
  {\bfseries 43} (1979) 1566--1570}.

\bibitem{Minkowski:1977sc}
P.~Minkowski, \emph{{$\mu \to e\gamma$ at a Rate of One Out of $10^{9}$ Muon
  Decays?}}, \href{https://doi.org/10.1016/0370-2693(77)90435-X}{\emph{Phys.
  Lett.} {\bfseries B67} (1977) 421--428}.

\bibitem{Ramond:1979py}
P.~Ramond, \emph{{The Family Group in Grand Unified Theories}},  in
  \emph{{International Symposium on Fundamentals of Quantum Theory and Quantum
  Field Theory Palm Coast, Florida, February 25-March 2, 1979}}, pp.~265--280,
  1979, \href{https://arxiv.org/abs/hep-ph/9809459}{{\ttfamily
  hep-ph/9809459}}.

\bibitem{GellMann:1980vs}
M.~Gell-Mann, P.~Ramond and R.~Slansky, \emph{{Complex Spinors and Unified
  Theories}}, {\emph{Conf. Proc.} {\bfseries C790927} (1979) 315--321},
  [\href{https://arxiv.org/abs/1306.4669}{{\ttfamily 1306.4669}}].

\bibitem{Yanagida:1979as}
T.~Yanagida, \emph{{HORIZONTAL SYMMETRY AND MASSES OF NEUTRINOS}}, {\emph{Conf.
  Proc.} {\bfseries C7902131} (1979) 95--99}.

\bibitem{Mohapatra:1979ia}
R.~N. Mohapatra and G.~Senjanovic, \emph{{Neutrino Mass and Spontaneous Parity
  Violation}}, \href{https://doi.org/10.1103/PhysRevLett.44.912}{\emph{Phys.
  Rev. Lett.} {\bfseries 44} (1980) 912}.

\bibitem{Schechter:1980gr}
J.~Schechter and J.~W.~F. Valle, \emph{{Neutrino Masses in SU(2) $\times$ U(1)
  Theories}}, \href{https://doi.org/10.1103/PhysRevD.22.2227}{\emph{Phys. Rev.}
  {\bfseries D22} (1980) 2227}.

\bibitem{Schechter:1981cv}
J.~Schechter and J.~W.~F. Valle, \emph{{Neutrino Decay and Spontaneous
  Violation of Lepton Number}},
  \href{https://doi.org/10.1103/PhysRevD.25.774}{\emph{Phys. Rev.} {\bfseries
  D25} (1982) 774}.

\bibitem{Mohapatra:1986aw}
R.~N. Mohapatra, \emph{{Mechanism for Understanding Small Neutrino Mass in
  Superstring Theories}},
  \href{https://doi.org/10.1103/PhysRevLett.56.561}{\emph{Phys. Rev. Lett.}
  {\bfseries 56} (1986) 561--563}.

\bibitem{Mohapatra:1986bd}
R.~N. Mohapatra and J.~W.~F. Valle, \emph{{Neutrino Mass and Baryon Number
  Nonconservation in Superstring Models}},
  \href{https://doi.org/10.1103/PhysRevD.34.1642}{\emph{Phys. Rev.} {\bfseries
  D34} (1986) 1642}.

\bibitem{Bernabeu:1987gr}
J.~{Bernab\'eu}, A.~Santamaria, J.~Vidal, A.~Mendez and J.~W.~F. Valle,
  \emph{{Lepton Flavor Nonconservation at High-Energies in a Superstring
  Inspired Standard Model}},
  \href{https://doi.org/10.1016/0370-2693(87)91100-2}{\emph{Phys. Lett.}
  {\bfseries B187} (1987) 303}.

\bibitem{Pilaftsis:1991ug}
A.~Pilaftsis, \emph{{Radiatively induced neutrino masses and large Higgs
  neutrino couplings in the standard model with Majorana fields}},
  \href{https://doi.org/10.1007/BF01482590}{\emph{Z. Phys.} {\bfseries C55}
  (1992) 275--282}, [\href{https://arxiv.org/abs/hep-ph/9901206}{{\ttfamily
  hep-ph/9901206}}].

\bibitem{Ilakovac:1994kj}
A.~Ilakovac and A.~Pilaftsis, \emph{{Flavor violating charged lepton decays in
  seesaw-type models}},
  \href{https://doi.org/10.1016/0550-3213(94)00567-X}{\emph{Nucl. Phys.}
  {\bfseries B437} (1995) 491},
  [\href{https://arxiv.org/abs/hep-ph/9403398}{{\ttfamily hep-ph/9403398}}].

\bibitem{Akhmedov:1995ip}
E.~K. Akhmedov, M.~Lindner, E.~Schnapka and J.~W.~F. Valle, \emph{{Left-right
  symmetry breaking in NJL approach}},
  \href{https://doi.org/10.1016/0370-2693(95)01504-3}{\emph{Phys. Lett.}
  {\bfseries B368} (1996) 270--280},
  [\href{https://arxiv.org/abs/hep-ph/9507275}{{\ttfamily hep-ph/9507275}}].

\bibitem{Akhmedov:1995vm}
E.~K. Akhmedov, M.~Lindner, E.~Schnapka and J.~W.~F. Valle, \emph{{Dynamical
  left-right symmetry breaking}},
  \href{https://doi.org/10.1103/PhysRevD.53.2752}{\emph{Phys. Rev.} {\bfseries
  D53} (1996) 2752--2780},
  [\href{https://arxiv.org/abs/hep-ph/9509255}{{\ttfamily hep-ph/9509255}}].

\bibitem{Barr:2003nn}
S.~M. Barr, \emph{{A Different seesaw formula for neutrino masses}},
  \href{https://doi.org/10.1103/PhysRevLett.92.101601}{\emph{Phys. Rev. Lett.}
  {\bfseries 92} (2004) 101601},
  [\href{https://arxiv.org/abs/hep-ph/0309152}{{\ttfamily hep-ph/0309152}}].

\bibitem{Malinsky:2005bi}
M.~{Malinsk\'y}, J.~C. {Rom\~{a}o} and J.~W.~F. Valle, \emph{{Novel
  supersymmetric SO(10) seesaw mechanism}},
  \href{https://doi.org/10.1103/PhysRevLett.95.161801}{\emph{Phys. Rev. Lett.}
  {\bfseries 95} (2005) 161801},
  [\href{https://arxiv.org/abs/hep-ph/0506296}{{\ttfamily hep-ph/0506296}}].

\bibitem{Branco:1988ex}
G.~C. Branco, W.~Grimus and L.~Lavoura, \emph{{The Seesaw Mechanism in the
  Presence of a Conserved Lepton Number}},
  \href{https://doi.org/10.1016/0550-3213(89)90304-0}{\emph{Nucl. Phys.}
  {\bfseries B312} (1989) 492--508}.

\bibitem{Buchmuller:1991tu}
W.~Buchmuller and C.~Greub, \emph{{Heavy Majorana neutrinos in electron -
  positron and electron - proton collisions}},
  \href{https://doi.org/10.1016/0550-3213(91)80024-G}{\emph{Nucl. Phys.}
  {\bfseries B363} (1991) 345--368}.

\bibitem{Asaka:2005an}
T.~Asaka, S.~Blanchet and M.~Shaposhnikov, \emph{{The nuMSM, dark matter and
  neutrino masses}},
  \href{https://doi.org/10.1016/j.physletb.2005.09.070}{\emph{Phys. Lett.}
  {\bfseries B631} (2005) 151--156},
  [\href{https://arxiv.org/abs/hep-ph/0503065}{{\ttfamily hep-ph/0503065}}].

\bibitem{Asaka:2005pn}
T.~Asaka and M.~Shaposhnikov, \emph{{The nuMSM, dark matter and baryon
  asymmetry of the universe}},
  \href{https://doi.org/10.1016/j.physletb.2005.06.020}{\emph{Phys. Lett.}
  {\bfseries B620} (2005) 17--26},
  [\href{https://arxiv.org/abs/hep-ph/0505013}{{\ttfamily hep-ph/0505013}}].

\bibitem{Chao:2009ef}
W.~Chao, Z.-g. Si, Y.-j. Zheng and S.~Zhou, \emph{{Testing the Realistic Seesaw
  Model with Two Heavy Majorana Neutrinos at the CERN Large Hadron Collider}},
  \href{https://doi.org/10.1016/j.physletb.2009.11.059}{\emph{Phys. Lett.}
  {\bfseries B683} (2010) 26--32},
  [\href{https://arxiv.org/abs/0907.0935}{{\ttfamily 0907.0935}}].

\bibitem{Chattopadhyay:2017zvs}
P.~Chattopadhyay and K.~M. Patel, \emph{{Discrete symmetries for electroweak
  natural type-I seesaw mechanism}},
  \href{https://doi.org/10.1016/j.nuclphysb.2017.06.008}{\emph{Nucl. Phys.}
  {\bfseries B921} (2017) 487--506},
  [\href{https://arxiv.org/abs/1703.09541}{{\ttfamily 1703.09541}}].

\bibitem{Adhikari:2010yt}
R.~Adhikari and A.~Raychaudhuri, \emph{{Light neutrinos from massless texture
  and below TeV seesaw scale}},
  \href{https://doi.org/10.1103/PhysRevD.84.033002}{\emph{Phys. Rev.}
  {\bfseries D84} (2011) 033002},
  [\href{https://arxiv.org/abs/1004.5111}{{\ttfamily 1004.5111}}].

\bibitem{Dev:2012sg}
P.~S.~B. Dev and A.~Pilaftsis, \emph{{Minimal Radiative Neutrino Mass Mechanism
  for Inverse Seesaw Models}},
  \href{https://doi.org/10.1103/PhysRevD.86.113001}{\emph{Phys. Rev.}
  {\bfseries D86} (2012) 113001},
  [\href{https://arxiv.org/abs/1209.4051}{{\ttfamily 1209.4051}}].

\bibitem{Kang:2006sn}
S.~K. Kang and C.~S. Kim, \emph{{Extended double seesaw model for neutrino mass
  spectrum and low scale leptogenesis}},
  \href{https://doi.org/10.1016/j.physletb.2006.12.071}{\emph{Phys. Lett.}
  {\bfseries B646} (2007) 248--252},
  [\href{https://arxiv.org/abs/hep-ph/0607072}{{\ttfamily hep-ph/0607072}}].

\bibitem{He:2009ua}
X.-G. He, S.~Oh, J.~Tandean and C.-C. Wen, \emph{{Large Mixing of Light and
  Heavy Neutrinos in Seesaw Models and the LHC}},
  \href{https://doi.org/10.1103/PhysRevD.80.073012}{\emph{Phys. Rev.}
  {\bfseries D80} (2009) 073012},
  [\href{https://arxiv.org/abs/0907.1607}{{\ttfamily 0907.1607}}].

\bibitem{Ibarra:2010xw}
A.~Ibarra, E.~Molinaro and S.~T. Petcov, \emph{{TeV Scale See-Saw Mechanisms of
  Neutrino Mass Generation, the Majorana Nature of the Heavy Singlet Neutrinos
  and $(\beta\beta)_{0\nu}$-Decay}},
  \href{https://doi.org/10.1007/JHEP09(2010)108}{\emph{JHEP} {\bfseries 09}
  (2010) 108}, [\href{https://arxiv.org/abs/1007.2378}{{\ttfamily 1007.2378}}].

\bibitem{Haba:2011pe}
N.~Haba, T.~Horita, K.~Kaneta and Y.~Mimura, \emph{{TeV-scale seesaw with
  non-negligible left-right neutrino mixings}},
  \href{https://arxiv.org/abs/1110.2252}{{\ttfamily 1110.2252}}.

\bibitem{Ma:2009du}
E.~Ma, \emph{{Deciphering the Seesaw Nature of Neutrino Mass from Unitarity
  Violation}}, \href{https://doi.org/10.1142/S0217732309031776}{\emph{Mod.
  Phys. Lett.} {\bfseries A24} (2009) 2161--2165},
  [\href{https://arxiv.org/abs/0904.1580}{{\ttfamily 0904.1580}}].

\bibitem{Wyler:1982dd}
D.~Wyler and L.~Wolfenstein, \emph{{Massless Neutrinos in Left-Right Symmetric
  Models}}, \href{https://doi.org/10.1016/0550-3213(83)90482-0}{\emph{Nucl.
  Phys.} {\bfseries B218} (1983) 205--214}.

\bibitem{Chikashige:1980ui}
Y.~Chikashige, R.~N. Mohapatra and R.~D. Peccei, \emph{{Are There Real
  Goldstone Bosons Associated with Broken Lepton Number?}},
  \href{https://doi.org/10.1016/0370-2693(81)90011-3}{\emph{Phys. Lett.}
  {\bfseries 98B} (1981) 265--268}.

\bibitem{Gelmini:1980re}
G.~B. Gelmini and M.~Roncadelli, \emph{{Left-Handed Neutrino Mass Scale and
  Spontaneously Broken Lepton Number}},
  \href{https://doi.org/10.1016/0370-2693(81)90559-1}{\emph{Phys. Lett.}
  {\bfseries 99B} (1981) 411--415}.

\bibitem{Leung:1983ti}
C.~N. Leung and S.~T. Petcov, \emph{{A Comment on the Coexistence of Dirac and
  Majorana Massive Neutrinos}},
  \href{https://doi.org/10.1016/0370-2693(83)91326-6}{\emph{Phys. Lett.}
  {\bfseries 125B} (1983) 461--466}.

\bibitem{Petcov:1984nz}
S.~T. Petcov and S.~T. Toshev, \emph{{Conservation of Lepton Charges, Massive
  Majorana and Massless Neutrinos}},
  \href{https://doi.org/10.1016/0370-2693(84)90829-3}{\emph{Phys. Lett.}
  {\bfseries 143B} (1984) 175--178}.

\bibitem{Chen:2011de}
M.-C. Chen and J.~Huang, \emph{{TeV scale models of neutrino masses and their
  phenomenology}}, \href{https://doi.org/10.1142/S0217732311035985}{\emph{Mod.
  Phys. Lett. A} {\bfseries 26} (2011) 1147--1167},
  [\href{https://arxiv.org/abs/1105.3188}{{\ttfamily 1105.3188}}].

\bibitem{Boucenna:2014zba}
S.~M. Boucenna, S.~Morisi and J.~W.~F. Valle, \emph{{The low-scale approach to
  neutrino masses}},  \href{https://arxiv.org/abs/1404.3751}{{\ttfamily
  1404.3751}}.

\bibitem{Aad:2015xaa}
{\scshape ATLAS} collaboration, G.~Aad et~al., \emph{{Search for heavy Majorana
  neutrinos with the ATLAS detector in pp collisions at $ \sqrt{s}=8 $ TeV}},
  \href{https://doi.org/10.1007/JHEP07(2015)162}{\emph{JHEP} {\bfseries 07}
  (2015) 162}, [\href{https://arxiv.org/abs/1506.06020}{{\ttfamily
  1506.06020}}].

\bibitem{Khachatryan:2016olu}
{\scshape CMS} collaboration, V.~Khachatryan et~al., \emph{{Search for heavy
  Majorana neutrinos in e$^{±}$e$^{±}$+ jets and e$^{±}$ $\mu^{±}$+ jets
  events in proton-proton collisions at $ \sqrt{s}=8 $ TeV}},
  \href{https://doi.org/10.1007/JHEP04(2016)169}{\emph{JHEP} {\bfseries 04}
  (2016) 169}, [\href{https://arxiv.org/abs/1603.02248}{{\ttfamily
  1603.02248}}].

\bibitem{Aaij:2014aba}
{\scshape LHCb} collaboration, R.~Aaij et~al., \emph{{Search for Majorana
  neutrinos in $B^- \to \pi^+\mu^-\mu^-$ decays}},
  \href{https://doi.org/10.1103/PhysRevLett.112.131802}{\emph{Phys. Rev. Lett.}
  {\bfseries 112} (2014) 131802},
  [\href{https://arxiv.org/abs/1401.5361}{{\ttfamily 1401.5361}}].

\bibitem{Shuve:2016muy}
B.~Shuve and M.~E. Peskin, \emph{{Revision of the LHCb Limit on Majorana
  Neutrinos}}, \href{https://doi.org/10.1103/PhysRevD.94.113007}{\emph{Phys.
  Rev.} {\bfseries D94} (2016) 113007},
  [\href{https://arxiv.org/abs/1607.04258}{{\ttfamily 1607.04258}}].

\bibitem{CERNNA48/2:2016tdo}
{\scshape NA48/2} collaboration, J.~R. Batley et~al., \emph{{Searches for
  lepton number violation and resonances in $K^{\pm}\to\pi\mu\mu$ decays}},
  \href{https://doi.org/10.1016/j.physletb.2017.03.029}{\emph{Phys. Lett.}
  {\bfseries B769} (2017) 67--76},
  [\href{https://arxiv.org/abs/1612.04723}{{\ttfamily 1612.04723}}].

\bibitem{Kersten:2007vk}
J.~Kersten and A.~{\relax Yu}. Smirnov, \emph{{Right-Handed Neutrinos at CERN
  LHC and the Mechanism of Neutrino Mass Generation}},
  \href{https://doi.org/10.1103/PhysRevD.76.073005}{\emph{Phys. Rev.}
  {\bfseries D76} (2007) 073005},
  [\href{https://arxiv.org/abs/0705.3221}{{\ttfamily 0705.3221}}].

\bibitem{Ingelman:1993ve}
G.~Ingelman and J.~Rathsman, \emph{{Heavy Majorana neutrinos at e p
  colliders}}, \href{https://doi.org/10.1007/BF01474620}{\emph{Z. Phys.}
  {\bfseries C60} (1993) 243--254}.

\bibitem{Gluza:2002vs}
J.~Gluza, \emph{{On teraelectronvolt Majorana neutrinos}}, {\emph{Acta Phys.
  Polon.} {\bfseries B33} (2002) 1735--1746},
  [\href{https://arxiv.org/abs/hep-ph/0201002}{{\ttfamily hep-ph/0201002}}].

\bibitem{Grimus:2000vj}
W.~Grimus and L.~Lavoura, \emph{{The Seesaw mechanism at arbitrary order:
  Disentangling the small scale from the large scale}},
  \href{https://doi.org/10.1088/1126-6708/2000/11/042}{\emph{JHEP} {\bfseries
  11} (2000) 042}, [\href{https://arxiv.org/abs/hep-ph/0008179}{{\ttfamily
  hep-ph/0008179}}].

\bibitem{AristizabalSierra:2011mn}
D.~Aristizabal~Sierra and C.~E. Yaguna, \emph{{On the importance of the 1-loop
  finite corrections to seesaw neutrino masses}},
  \href{https://doi.org/10.1007/JHEP08(2011)013}{\emph{JHEP} {\bfseries 08}
  (2011) 013}, [\href{https://arxiv.org/abs/1106.3587}{{\ttfamily 1106.3587}}].

\bibitem{Lopez-Pavon:2015cga}
J.~Lopez-Pavon, E.~Molinaro and S.~T. Petcov, \emph{{Radiative Corrections to
  Light Neutrino Masses in Low Scale Type I Seesaw Scenarios and Neutrinoless
  Double Beta Decay}},
  \href{https://doi.org/10.1007/JHEP11(2015)030}{\emph{JHEP} {\bfseries 11}
  (2015) 030}, [\href{https://arxiv.org/abs/1506.05296}{{\ttfamily
  1506.05296}}].

\bibitem{Korner:1992zk}
J.~G. Korner, A.~Pilaftsis and K.~Schilcher, \emph{{Leptonic CP asymmetries in
  flavor changing H0 decays}},
  \href{https://doi.org/10.1103/PhysRevD.47.1080}{\emph{Phys. Rev.} {\bfseries
  D47} (1993) 1080--1086},
  [\href{https://arxiv.org/abs/hep-ph/9301289}{{\ttfamily hep-ph/9301289}}].

\bibitem{LopezPavon:2012zg}
J.~Lopez-Pavon, S.~Pascoli and C.-f. Wong, \emph{{Can heavy neutrinos dominate
  neutrinoless double beta decay?}},
  \href{https://doi.org/10.1103/PhysRevD.87.093007}{\emph{Phys. Rev.}
  {\bfseries D87} (2013) 093007},
  [\href{https://arxiv.org/abs/1209.5342}{{\ttfamily 1209.5342}}].

\bibitem{Pascoli:2013fiz}
S.~Pascoli, M.~Mitra and S.~Wong, \emph{{Effect of cancellation in neutrinoless
  double beta decay}},
  \href{https://doi.org/10.1103/PhysRevD.90.093005}{\emph{Phys. Rev.}
  {\bfseries D90} (2014) 093005},
  [\href{https://arxiv.org/abs/1310.6218}{{\ttfamily 1310.6218}}].

\bibitem{Haba:2016lxc}
N.~Haba, H.~Ishida and Y.~Yamaguchi, \emph{{Naturalness and lepton
  number/flavor violation in inverse seesaw models}},
  \href{https://doi.org/10.1007/JHEP11(2016)003}{\emph{JHEP} {\bfseries 11}
  (2016) 003}, [\href{https://arxiv.org/abs/1608.07447}{{\ttfamily
  1608.07447}}].

\bibitem{Hernandez:2016kel}
P.~Hern\'andez, M.~Kekic, J.~L\'opez-Pav\'on, J.~Racker and J.~Salvado,
  \emph{{Testable Baryogenesis in Seesaw Models}},
  \href{https://doi.org/10.1007/JHEP08(2016)157}{\emph{JHEP} {\bfseries 08}
  (2016) 157}, [\href{https://arxiv.org/abs/1606.06719}{{\ttfamily
  1606.06719}}].

\bibitem{Drewes:2016lqo}
M.~Drewes and S.~Eijima, \emph{{Neutrinoless double $\beta$ decay and low scale
  leptogenesis}},
  \href{https://doi.org/10.1016/j.physletb.2016.09.054}{\emph{Phys. Lett.}
  {\bfseries B763} (2016) 72--79},
  [\href{https://arxiv.org/abs/1606.06221}{{\ttfamily 1606.06221}}].

\bibitem{Fernandez-Martinez:2016lgt}
E.~Fernandez-Martinez, J.~Hernandez-Garcia and J.~Lopez-Pavon, \emph{{Global
  constraints on heavy neutrino mixing}},
  \href{https://doi.org/10.1007/JHEP08(2016)033}{\emph{JHEP} {\bfseries 08}
  (2016) 033}, [\href{https://arxiv.org/abs/1605.08774}{{\ttfamily
  1605.08774}}].

\bibitem{Kraus:2004zw}
C.~Kraus et~al., \emph{{Final results from phase II of the Mainz neutrino mass
  search in tritium beta decay}},
  \href{https://doi.org/10.1140/epjc/s2005-02139-7}{\emph{Eur. Phys. J.}
  {\bfseries C40} (2005) 447--468},
  [\href{https://arxiv.org/abs/hep-ex/0412056}{{\ttfamily hep-ex/0412056}}].

\bibitem{Aseev:2011dq}
{\scshape Troitsk} collaboration, V.~N. Aseev et~al., \emph{{An upper limit on
  electron antineutrino mass from Troitsk experiment}},
  \href{https://doi.org/10.1103/PhysRevD.84.112003}{\emph{Phys. Rev.}
  {\bfseries D84} (2011) 112003},
  [\href{https://arxiv.org/abs/1108.5034}{{\ttfamily 1108.5034}}].

\bibitem{Ade:2015xua}
{\scshape Planck} collaboration, P.~A.~R. Ade et~al., \emph{{Planck 2015
  results. XIII. Cosmological parameters}},
  \href{https://doi.org/10.1051/0004-6361/201525830}{\emph{Astron. Astrophys.}
  {\bfseries 594} (2016) A13},
  [\href{https://arxiv.org/abs/1502.01589}{{\ttfamily 1502.01589}}].

\end{thebibliography}\endgroup
\bibliographystyle{JHEP}

\end{document}